# Dual-comb correlation spectroscopy reveals laser dynamics

Xiuxiu Zhang[1]†, Zhuoren Wan[1]†, Yuling Sheng[1], Ming Yan[1,2]*, Yuan Chen[1], Zijian Wang[1], Zhaoyang Wen[1,2], Min Li[3]*, and Heping Zeng[1,2,4]*

[1]State Key Laboratory of Precision Spectroscopy, and Hainan Institute, East China Normal University, Shanghai, China

[2]Chongqing Key Laboratory of Precision Optics, Chongqing Institute of East China Normal University, Chongqing 401120, China

[3]School of Optical Electrical and Computer Engineering, University of Shanghai for Science and Technology, Shanghai 200093, China

[4]Jinan Institute of Quantum Technology, Jinan, Shandong 250101, China

† These authors contribute equally.

* Corresponding author: *myan@lps.ecnu.edu.cn, minli1220@usst.edu.cn, hpzeng@phy.ecnu.edu.cn*

**Abstract:**

Laser dynamics underpin a broad range of modern photonic technologies and continue to reveal rich nonlinear behaviors. However, existing spectroscopic tools—most notably time-stretched dispersive Fourier transform spectroscopy (TS-DFT)—remain limited in spectral resolution, accuracy, and their ability to capture continuous waveforms and complex field dynamics. Here, we introduce dual-comb correlation spectroscopy (DCCS) as a powerful approach for resolving fast and intricate laser behaviors that are inaccessible to TS-DFT and conventional spectrometers. By correlating two sequences of heterodyne spectra produced by mixing a test laser with a pair of optical combs, DCCS enables rapid (e.g., 1 µs) and high-resolution (0.08 pm) spectral retrieval over broad optical bandwidths. Leveraging these capabilities, we reveal mode-hopping and mode-competition dynamics in continuous-wave lasers, as well as the buildup process of a mode-locked laser. These results establish DCCS as a versatile and complementary tool to TS-DFT for exploring transient, broadband, and previously unresolvable behaviors in lasers and other time-evolving optical systems.

# Introduction

Understanding the dynamic behavior of lasers, such as noise build-up, mode competition, and pulse formation, is essential for advancing photonic technologies ranging from high-speed communication [1] to precision metrology [2]. However, conventional Fourier-transform spectrometers and camera-based dispersive spectrometers are too slow to capture these phenomena [3]. Under this circumstance, time-stretch dispersive Fourier transform spectroscopy (TS-DFT) has emerged as a powerful method for probing soliton dynamics in real time [4-6]. By mapping spectral information into the time domain using chromatic dispersion, TS-DFT has enabled real-time monitoring of transient phenomena such as soliton formation [7], optical rogue waves [8], and Q-switching instabilities [9] in mode-locked lasers. Despite its success, TS-DFT inherently measures only the spectral intensity of coherent pulses with a limited spectral resolution (e.g., 0.01-1 nm [10-12]), lacking direct access to phase and frequency-mode information—quantities that are essential for fully characterizing the complex field evolution of mode-locked and continuous-wave (CW) lasers [13, 14].

Optical frequency combs, with their precisely spaced and phase-coherent spectral lines, have opened new frontiers in both spectroscopic and temporal resolution [15, 16]. Among the most impactful applications is dual-comb spectroscopy (DCS), which leverages the interference between two combs with slightly detuned repetition rates to perform high-resolution, broadband measurements without mechanical scanning [17-20]. DCS has revolutionized fields such as precision spectroscopy [21], distance metrology [22], and hyperspectral imaging [23]. However, conventional DCS is an active spectroscopic technique, wherein the combs directly interrogate a sample—limiting its applicability to passive or spontaneous optical phenomena that cannot be externally excited [24].

To address this limitation, dual-comb correlation spectroscopy (DCCS) has emerged as a passive measurement technique [25-27]. In DCCS, two asynchronous combs independently sample an external light source. By computing the temporal cross-correlation of the acquired signals, the system reconstructs the source's spectral and phase contents. This approach has been successfully demonstrated in the microwave domain [26] and in the spectral reconstruction of natural incoherent light [27], though it has so far been restricted to static or quasi-static measurements.

In this work, we extend DCCS for the first time to the investigation of laser dynamics, demonstrating its previously unexplored ability to retrieve the complex behaviors—including intensity, frequency modes, and phase—of a laser during its startup process. Using a high-repetition-rate dual-comb system, we capture rapid transitions from amplified spontaneous emission (ASE) to the emergence of a CW background and the eventual formation of coherent pulses. These results offer insights into transient processes that are inaccessible with direct TS-DFT or conventional spectrometers.

# Results

## Principles and experimental schematics

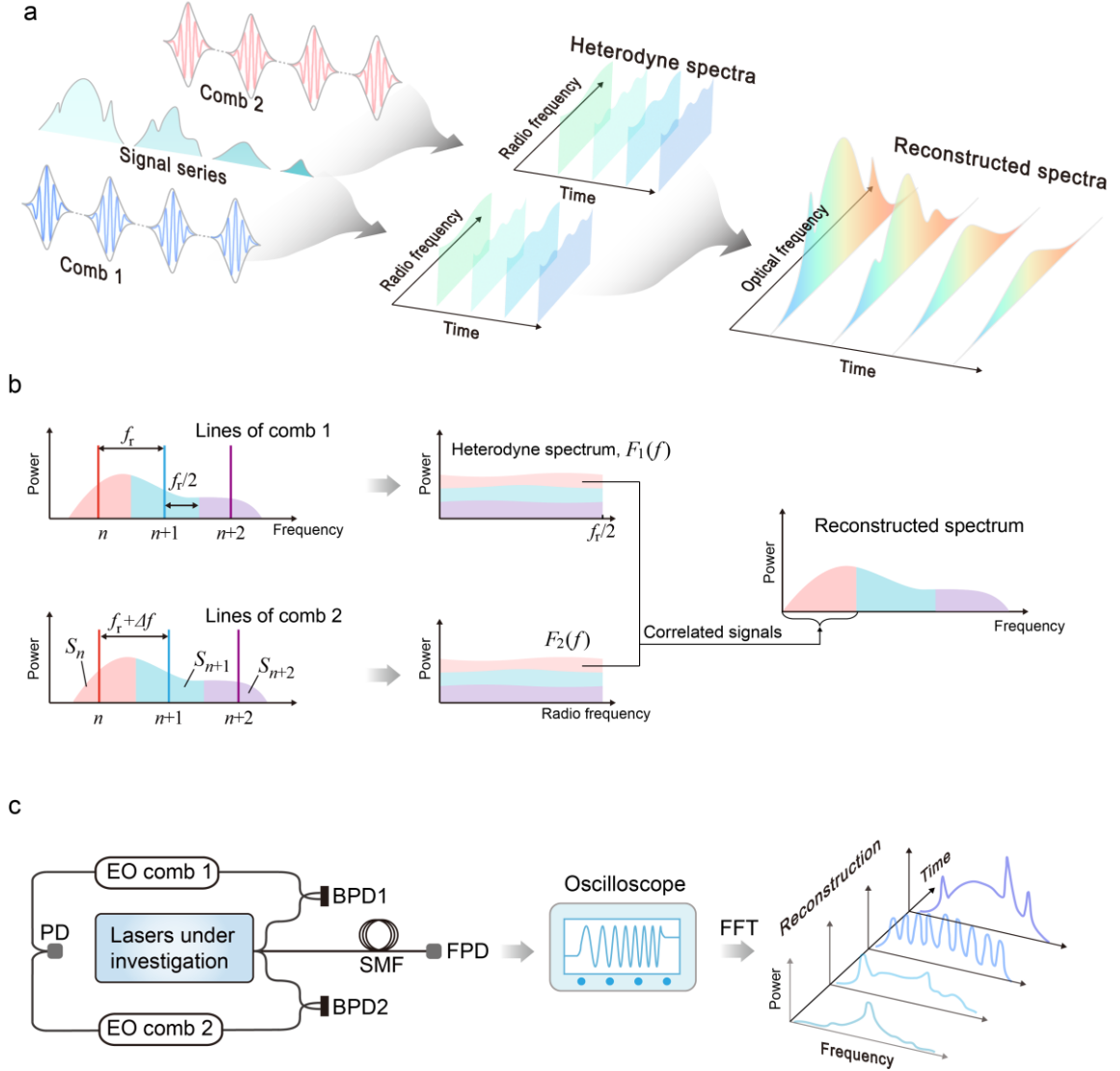

**Fig. 1 | Dual-comb correlation spectroscopy (DCCS) revealing laser dynamics. a,** Schematic of the DCCS concept. **b,** Basic principle. Here, $n$ denotes the comb mode index, and $S_n$ represents the $n$-th sub-spectrum of the test emission. We assmue the two comb spectra are identical and broader than the measured spectrum. **c,** Experimental setup. PD, photodetector; FPD, fast photodetector; BPD, balanced photodetector; SMF, single-mode fiber; $f_{\text{rep}}$, repetition rate; $\Delta f$, repetition rate difference; FFT, Fourier-transformation.

As shown in Fig. 1a, DCCS utilizes two optical combs with slightly different repetition rates ($f_r$ and $f_r+\Delta f$), each heterodyning with the same signal under test. This process generates two correlated heterodyne spectra in the radio-frequency domain, which are then combined computationally to reconstruct the signal spectrum. Detailed mathematical descriptions of DCCS have been thoroughly elaborated in the literature [25-27]. Here, we briefly outline its core principles. As illustrated in Fig. 1b, the spectrum of the test light source, $S(f)$, can be regarded as a collection of sub-spectra, $S_n(f_n+f)$, each centered at a comb-line frequency $f_n$ and spanning from $f_n-f_r/2$ to $f_n+f_r/2$. When the test light is heterodyned with two combs on separate detectors, it produces two correlated heterodyne spectra, $F_1(f)$ and $F_2(f)$, defined within the frequency range 0 to $f_r/2$. Each heterodyne spectrum contains the superposition of beat signals arising from the overlap between the test spectrum and all comb lines. To isolate the heterodyne contribution associated with the $n$-th sub-spectrum, we compute the Fourier transform of the cross-correlation between $F_1(f)$ and $F_2(f)$:

$$C_n(f) \equiv \mathcal{F}_T \left[ F_2(f - n \cdot \Delta f, T) \, F_1^*(f, T) \right](-n \cdot \Delta f),$$

where $F_i(f, T)$ denotes the heterodyne spectra measured over the obversation time $T$, and $\Delta f$ is the repetition-rate difference between the two combs. The $n$-th sub-spectrum of the test source is then resonstructed as

$$S_n(f_n + f) = \langle C_n(f) \rangle \cdot A_n^{-1}$$

where $A_n$ is the amplitude of the beat note between the $n$-th comb lines of the two combs, and $\langle \cdot \rangle$ denotes averaging over repeated measurements. The full spectrum $S(f)$ is obtained by joining the sub-spectra $S_n(f_n + f)$ for $n=1, 2, ..., N$, where $N$ is the total number of comb lines in each comb.

When applied to the study of laser dynamics, DCCS offers several key advantages. (1) It can resolve longitudinal laser modes and phase with exceptionally high resolution, set primarily by the Fourier-transform time window. (2) Compared with laser heterodyne radiometry (LHR), in which the test source is heterodyned with a single-frequency CW laser, DCCS provides broadband spectral access and removes spectral ambiguities [26]. (3) It enables rapid measurements with temporal resolutions set by of $1/\Delta f$, which can be as short as a few microseconds (μs), or potentially down to the nanosecond (ns) level [28]. Combining these advantages, DCCS is expected to reveal laser dynamic processes that are otherwise 'invisible' to traditional techniques.

Figure 1c illustrates our experimental setup. We employed two electro-optic (EO) combs with $f_r$=25 GHz and $\Delta f$ =1 MHz, supporting data refresh times as short as 1 μs. Both combs were seeded by the same CW laser at 1550 nm (see Methods), ensuring an excellent mutual coherence with dual-comb beat-note linewidths below 1 Hz. The two combs were mixed on a photodetector to generate standard dual-comb signals for spectral calibration. For DCCS measurements, the laser under investigation was equally split into two paths and heterodyned with the two combs on separate balanced photodetectors (BPDs, 20 GHz bandwidth). Meanwhile, a small fraction of the test laser was analyzed using a TS-DFT system comprising a 10-km single-mode fiber (dispersion, –24 ps/nm/km) and a high-speed photodetector (40 GHz bandwidth). The outputs of all four detectors were simultaneously recorded by a four-channel high-speed oscilloscope (33 GHz bandwidth, 20 GHz sampling rate), and the acquired data were post-processed to reconstruct the test laser spectrum. Further experimental details are provided in Supplementary Fig. S1.

**Spectral reconstruction**

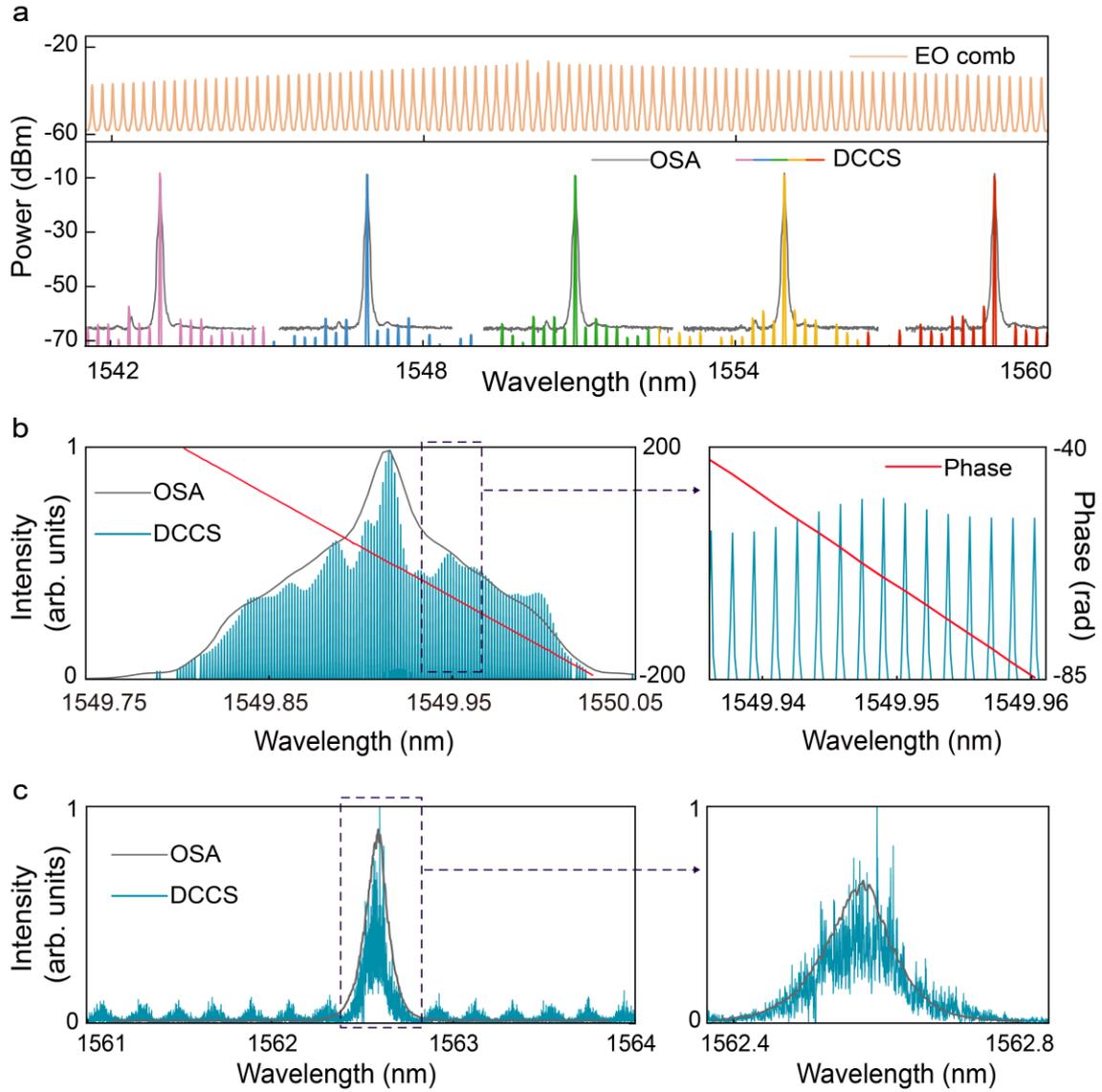

**Fig. 2 | Spectral results. a,** Spectral reconstructions of a wavelength-tunable continuous-wave (CW) laser compared with measurements from an optical spectrum analyzer (OSA). Shown above is the spectrum of a single electro-optic (EO) comb with a 25 GHz repetition rate. **b,** Reconstruction of a frequency-agile comb with a 200 MHz repetition rate, compared with OSA measurements. A zoomed-in view is shown on the right. **c,** Reconstructed ASE spectrum compared with the corresponding OSA measurement. Addtional spectral details are shown on the right.

First, we validated our system by measuring steady-state spectra of a wavelength-tunable CW laser, a frequency-agile comb [29], and an ASE source. Figure 2a shows reconstructed spectra (color curves) of the CW laser, tuned from 1542.7 nm to 1559.6 nm, within the EO comb spectral range (1539–1561 nm at -10 dB, orange curve). Our results match the data (grey curves) measured by a commercial optical spectral analyzer (OSA, resolution of 20 pm), while achieving superior 10 MHz (0.08 pm) resolution with 100 µs acquisition time per spectrum. The signal-to-noise ratio (SNR) exceeds 50 dB, defined as the ratio of peak signal to noise floor deviation. Unlike conventional heterodyne detection between a comb and a test CW laser of unknown frequency, which requires multiple measurements to determine the absolute frequency, DCCS achieves this in a single measurement.

Furthermore, DCCS has broadband reconstruction capability, enabling simultaneous measurements of multi-mode signals. Figure 2b illustrates measurements of a frequency-agile comb with ~130 evenly spaced modes (Methods). Compared to the OSA spectrum (20 pm resolution, red curve) with a 10 ms acquisition time, DCCS resolves the fine comb structure (0.16 pm resolution) in just 200 µs. Additionally, it provides phase information of the resolved modes (Fig. 2b, red line), setting it apart from conventional intensity-only spectrometers.

Advantageously, DCCS can measure both coherent and weakly-coherent (or even incoherent) radiation. For example, we measured ASE spectra from a fiber laser in a non-mode-locked state (Methods). The reconstructed spectra (blue), along with a steady-state spectrum (grey), are shown in Fig. 2c. The reconstruction used a 1 ms acquisition time and achieved a 30 MHz (or 0.24 pm) spectral resolution. A zoom-in view further highlights the finely resolved noisy spectral characteristics. These results demonstrate the rapid, high-resolution, and mode-resolvable capability of DCCS for characterizing light emission under test.

**Revealing CW laser dynamics**

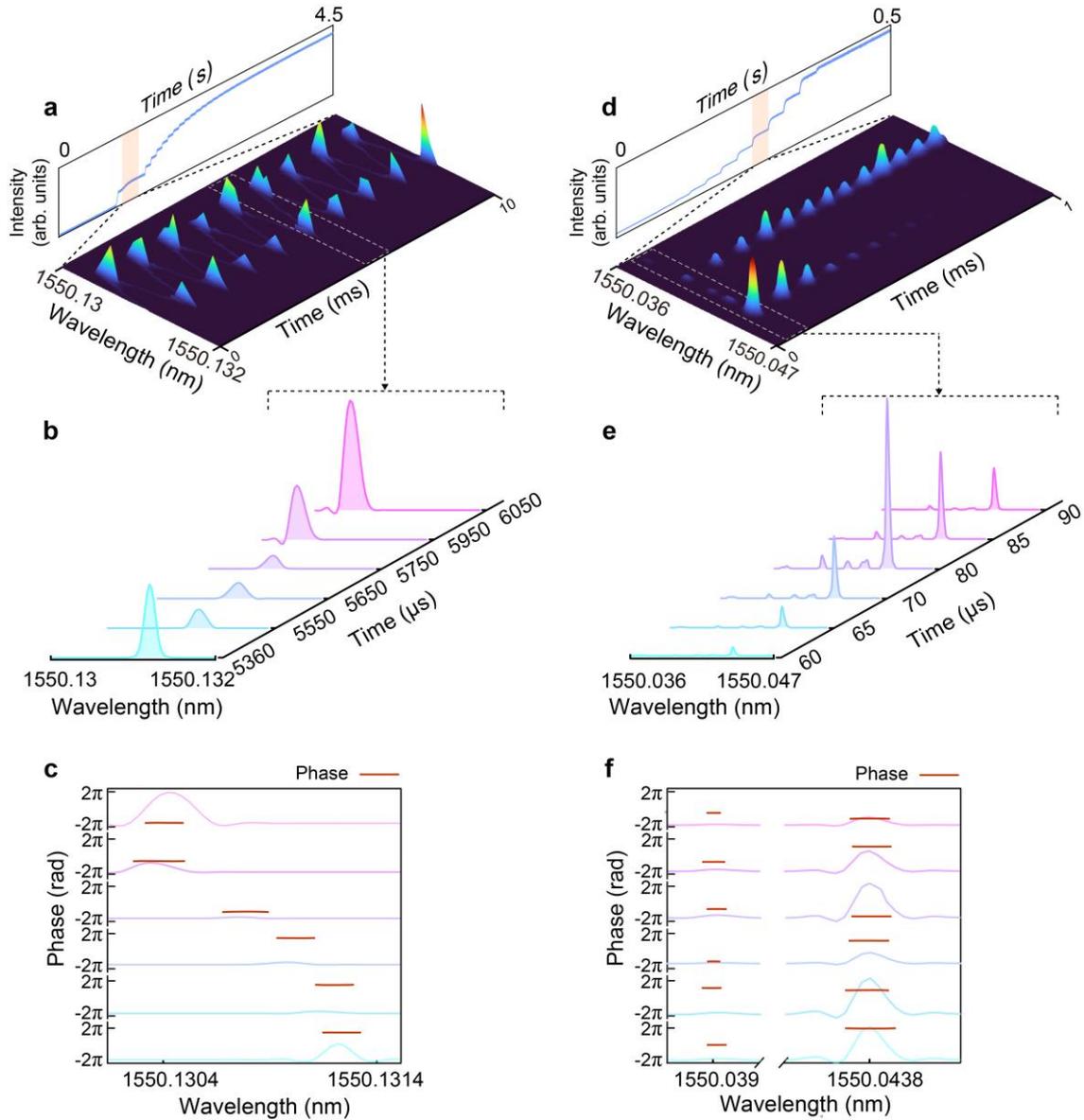

**Fig. 3 | Transient dynamics of CW lasers. a,** Mode hopping in a semiconductor CW laser, measured with a spectral resolution of 0.08 pm at a 100 kHz refresh rate. **b,** Zoom-in view of the spectra highlighted by the dashed box in **a**. **c**, Corresponding phase spectra of the resolved modes. **d,** Mode competition in a fiber-ring CW laser, measured at a 1 MHz refresh rate. **e,** Zoom-in view of the spectra highlighted by the dashed box in **d**. **f,** Reconstructed phase information corresponding to the displayed modes.

Next, we investigated transient behaviors in CW lasers, focusing on abrupt longitudinal-mode transitions (mode hopping) and multi-mode gain competition, as they critically affect laser stability and reliability. As a starting point, we measured an off-the-shelf semiconductor CW laser. When the pump current was switched on, the laser exhibited rapid, periodic mode hopping over a spectral span of 2 pm (≈250 MHz), as shown in Fig. 3a. Capturing such dynamics requires both high acquisition speed and fine spectral resolution—capabilities that remain challenging for conventional techniques. Also, because standard TS-DFT lacks the ability to resolve continuous waveforms spectrally, it recorded only the temporal power (intensity) variations (blue curve in Fig. 3a). In contrast, DCCS achieved transient detection with a 10 µs integration time and 0.08 pm (10 MHz) resolution. Examples of the measured frequency modes and their phases are shown in Fig. 3b and Fig. 3c, respectively. Two-dimensional maps of phase evolution are provided in Supplementary Fig. S2.

We note that while LHR can be used for optical frequency measurement when the test laser frequency is approximately known, it becomes unsuitable for multi-mode lasers due to frequency ambiguity and limited detection bandwidth. DCCS overcomes this limitation. To illustrate this, we built an FBG-based erbium-doped fiber ring laser emitting multiple wavelengths (see Methods). Using DCCS, we reconstructed the startup dynamics and revealed fast competition between two dominant modes (Fig. 3d) on a 10-µs time scale. Figures 3e and 3f show zoom-in views of the frequency modes and phases, respectively. Although the spectra are shown within a narrow wavelength window (0.011 nm), our measurements actually covered a broader range defined by the dual combs. Transient, broadband measurements (spanning 1547–1553 nm) of a multimode semiconductor laser are demonstrated in Supplementary Fig. S3.

**Reconstruction of start-up dynamics in mode-locked lasers**

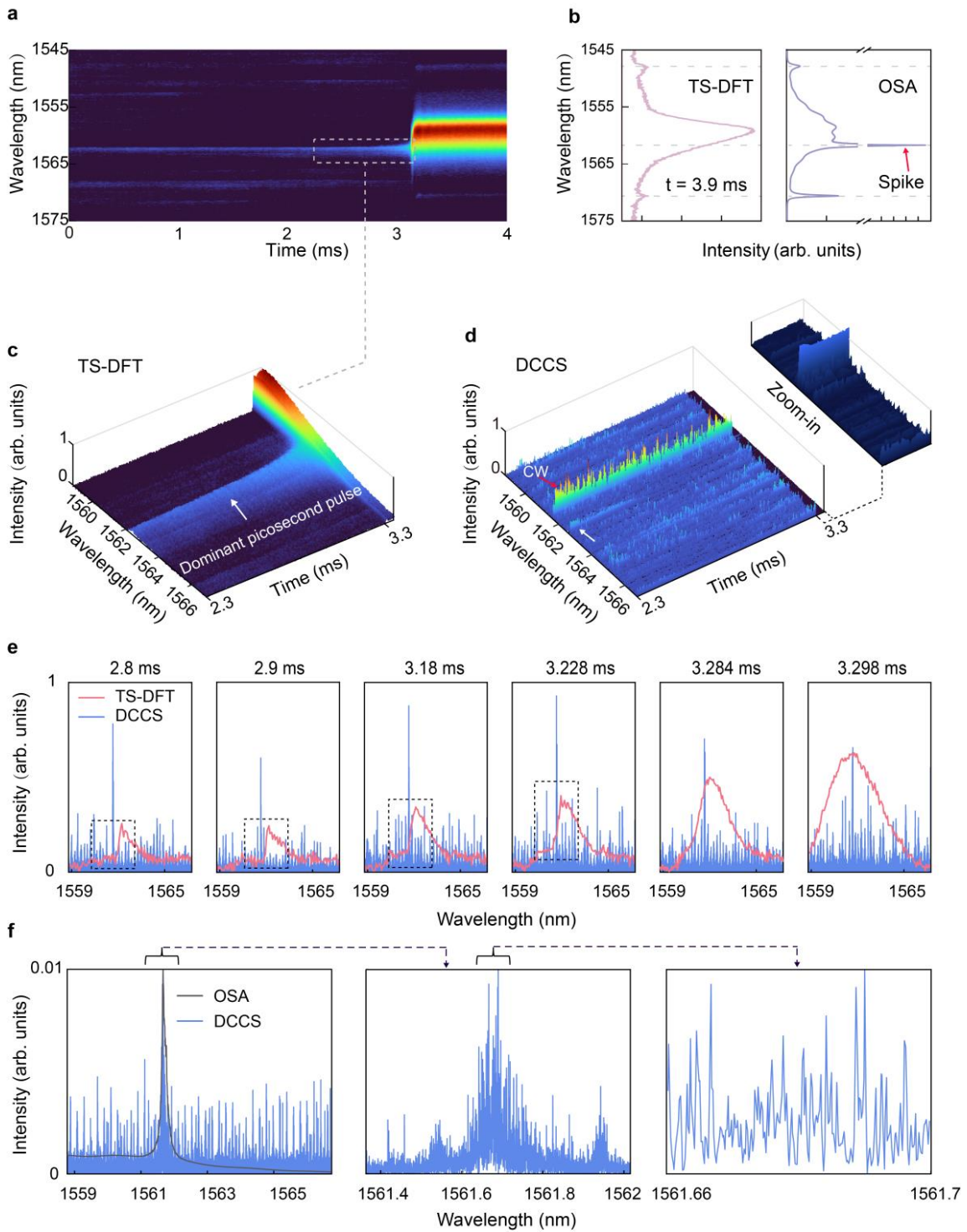

**Fig. 4 | Build-up dynamics of a mode-locked laser. a,** Time-stretched 2D spectrogram. **b,** Comparison between time-stretched disperisve Fourier transform (TS-DFT) spectra and steady-state spectra measured by an optical spectral analyzer (OSA). **c,** Zoom-in view of the dashed box region in **a**. **d,** Corresponding results obtained using dual-comb cross-correlation spectroscopy (DCCS). **e,** Sanpshots of the TS-DFT

and DCCS results. **f,** DCCS results (blue) compared with OSA measurements (gray) for a broadband mode-locked state.

Finally, we applied DCCS to investigate mode-locked laser dynamics. We constructed a nonlinear-polarization-rotation (NPR) mode-locked fiber laser (Supplementary Fig. S1), emitting pulses at a 21.7 MHz repetition rate with an average output power of 3 mW. The laser output was split into two paths: one (0.5 mW) directed to the TS-DFT system to generate a reference spectrum, and the other (2.5 mW) to the DCCS system for dual-comb heterodyne detection. Optical and electronic delays between the two paths were pre-calibrated to synchronize the TS-DFT and DCCS measurements.

Figure 4a shows a representative 2D TS-DFT spectrogram capturing the laser startup dynamics, i.e., the evolution from a narrowband pulsing state to a broadband mode-locked state. Broadband mode-locking emerges at an observation time of t=3.1 ms, with a central wavelength of 1560 nm and first-order Kelly sidebands located at 1547.85 nm and 1570.57 nm (Fig. 4b). In particular, we investigated both the transition stage (dashed box in Fig. 4a) and the subsequent mode-locked state, in which a strong spectral spike—normally attributed to a quasi-CW component—appears in the steady-state spectrum but is absent in the TS-DFT measurement (Fig. 4b).

We plotted the 3D TS-DFT and DCCS spectrograms (spanning 1558.8–1566.74 nm over a duration exceeding 1 ms) in Figs. 4c and 4d, respectively. The TS-DFT measurement reveals a narrowband spectral feature at 1562.8 nm (Fig. 4c), indicating the presence of a dominant picosecond pulse [30]. This agrees with the DCCS spectrogram, which shows a corresponding spectral component at the same wavelength (marked by a white arrow in Fig. 4d). Note that this weak heterodyne signal gradually fades because, as the pulse spectrum broadens, the power per spectral element decreases, leading to smaller heterodyne amplitudes that eventually sink below the noise floor. Nevertheless, we emphasize two relevant findings: (1) a strong CW component at 1561.7 nm (red arrow) appears in the DCCS measurement but is absent in the TS-DFT results; and (2) the TS-DFT spectral profile of the dominant pulse exhibits an asymmetric shape, as highlighted by the pink curves within the black dashed boxes in Fig. 4e. We attribute this asymmetric Fano-like resonance [31] to the interaction between the CW component (blue peak in Fig. 4e) and the dominant pulse. Numerical simulations (Supplementary Fig. S4) corroborate this mechanism and show excellent agreement with the experimental data.

After the laser entered the broadband mode-locked state, the CW component abruptly dropped (zoom-in view in Fig. 4d) and became spectrally noisy (Fig. 4f). This noisy, unstable signal is the origin of the spike observed by the OSA (in Fig. 4f and Fig. 4b). This finding is generally consistent with previous studies, which attributed the spike to a quasi-CW component [32]. We note that, in contrast to the OSA data in Fig. 4f—acquired with a 10 pm (1.2 GHz) resolution—DCCS attains a substantially finer resolution of 0.08 pm (10 MHz), thereby resolving the spectral structure hidden beneath the envelope.

## Discussion

DCCS offers excellent spectral resolving capability for measuring continuous waveforms and multi-mode signals, making it a unique and complementary tool to existing techniques, including TS-DFT, for probing fast-evolving laser dynamics. For example, the dual-comb measurements in Figs. 4d and 4e provide direct evidence of a CW component responsible for the Fano-like spectral profile observed in the TS-DFT results (Figs. 4c and 4e).

We also note that the temporal resolution of DCCS is currently limited to $1/\Delta f = 1$ μs, which is incompatible with TS-DFT operating at the pulse round-trip time of the mode-locked laser (i.e. 46 ns, corresponding to a 21.7 MHz repetition rate). This limitation can be overcome by using two microresonator combs for DCCS. Such micronized combs feature large line spacings (e.g., 100 GHz) and $\Delta f$ values on the order of 100 MHz [33-35], making them suitable for resolving dynamics in most mode-locked lasers [36, 37].

Nevertheless, owing to its ability to rapidly resolve complex field evolutions together with phase and frequency-mode information, DCCS is expected to advance studies of nonlinear pulse formation [38], soliton explosions [39], Q-switching instabilities [40], and cavity buildup dynamics [41]. In particular, we anticipate that integrating TS-DFT with DCCS will provide a comprehensive framework for exploring fundamental laser physics—including modulation instability [42], chaotic lasing behavior [43], and dissipative-structure dynamics [44]—by enabling direct identification of transient coherence waves, spectral fringes, and noise-driven fluctuations. These combined capabilities will be valuable for laser-stability enhancement, narrow-linewidth laser design, and advanced laser engineering.

In conclusion, we have demonstrated that DCCS provides rapid, high-resolution access to transient and complex laser dynamics. Experimentally, DCCS enabled the direct observation of mode hopping and multi-mode competition with both high spectral resolution and a broad detection bandwidth. In addition, by combining DCCS with TS-DFT, we unveiled previously unresolved features of the transient buildup dynamics in a mode-locked laser. By introducing this new diagnostic capability, our work advances fundamental studies of laser dynamics and offers valuable tools for future laser engineering.

## Author contributions

M. Y. and H. Z. conceived the idea. M. Y. and M. L. designed the experiments. X. Z., Z. W., and Y. S. conducted the experiment. Y. C. and Z.J. W. build the comb source. X. Z. and Z.Y. W. build the mode-locked laser and the single-mode laser. X. Z. and Z. W. drafted the manuscript and analyzed the data. M. Y., M. L., and H. Z. revised the manuscript. All authors provided comments and suggestions for improvements.

## Competing interests

The authors declare no competing interests.

## Data availability

The data that support the findings of this study are available from the corresponding author upon reasonable request.

## Author information

Correspondence and requests for materials should be addressed to M.Y. (myan@lps.ecnu.edu.cn), M. L. (minli1220@usst.edu.cn), or to H.Z. (hpzeng@phy.ecnu.edu.cn).